\documentclass[twocolumn,showpacs,preprintnumbers,amsmath,amssymb]{revtex4-1}
\usepackage{graphicx}
\usepackage{ulem}
\usepackage{epsfig}
\usepackage{amsmath}
\usepackage{amssymb}

\usepackage{color}
\usepackage[usenames,dvipsnames]{xcolor}

\begin{document}

\title{Numerical study of the depinning transition of a ferromagnetic magnetic domain wall in films}

\author{Bin Xi$^1$, Meng-Bo Luo$^2$, Valerii M. Vinokur$^3$ \& Xiao Hu$^1$}

\affiliation{
          $^1${International Center for Materials Nanoarchitectonics (WPI-MANA), National Institute for Materials Science, Tsukuba 305-0044, Japan}\\
          $^2${Department of Physics, Zhejiang University, Hangzhou 310027, China}\\
          $^3${Materials Science Division, Argonne National Laboratory, Argonne, Illinois 60439, USA}
 }

\begin{abstract}
We report first principle numerical study of domain wall (DW) depinning in two-dimensional magnetic film,
which is modeled by 2D random-field Ising system with the dipole-dipole interaction.
We observe nonconventional activation-type motion of DW and reveal its fractal structure of DW near the depinning transition.
We determine scaling functions describing critical dynamics near the transition and obtain universal exponents establishing
connection between thermal softening of pinning potential and critical dynamics.
We observe that tuning the strength of the dipole-dipole interaction switches DW dynamics between two different universality
classes corresponding to two distinct dynamic regimes, motion in the random potential and that in the random force.
\end{abstract}

\date{\today}

\maketitle

\noindent \textit{Introduction --}
Motion of domain walls in magnetic nanowires and films is a key
component of operation of any magnetic memory and logic
device\cite{Parkin:2008,Hayashi:2008,Allwood:2005}.
To a great extent DW dynamics is governed by pinning-depinning processes
which control the operational speed and power consumption
of a device and thus play central role in device performance\cite{Miron:2011,Kim:2013}.
There has been remarkable progress in description of pinned DW dynamics based
mostly on the elastic manifold model in a random environment\cite{Ioffe:1987}.
A key property of such a system is that at zero temperature it experiences
the \textit{dynamic} phase transition (depinning transition):
At small external drives, $F\leqslant F_c$, where $F_c$ is the critical pinning force,
DW is immobilized (pinned) by disorder, whereas at $F>F_c$ it acquires a finite
velocity $v$.  The threshold depinning force $F_c$ is a critical point in a sense
that at $F\gtrsim F_c$, the velocity exhibits critical behavior
$v\sim(F-F_c)^{\beta}$, as was proposed by Fisher\cite{Fisher:1983} in the
context of depinning of charge density waves.
At finite temperatures the velocity is always finite,
and at $F\ll F_c$ the domain wall exhibits highly nonlinear glassy response,
so-called \textit{creep} dynamics,
with $v\propto\exp(-\mathrm{const}/F^{\mu})$\cite{Ioffe:1987}.
The depinning transition gets rounded and acquires a meaning of the
intermediate region separating the low force creep dynamics from
the asymptotic linear response $v\propto F$ at $F\gg F_c$.
The critical depinning behavior has to include temperature dependence
and was conjectured to be of the form:
$v\sim\Psi \left[\left(F/F_c-1\right )/T^{\eta}\right]$ \cite{PokrNattVin:2001,Glatz:2003}.

All the above results were obtained within the elastic manifold approach,
where the DW was modeled as an elastic membrane (in 3D) or an elastic string,
if we discuss one-dimensional DW in a magnetic film.
However fundamental and successful, this description misses important processes that
may become essential for the DW dynamics at elevated temperatures.
Domain wall is multi-valued, so ahead boundaries can merge with the `main' interface behind.

This poses a challenge of developing first principle approach starting from the
microscopic model that captures basic physics of the magnetic system.
Taking up upon this challenge we reveal the fractal structure of the
domain wall and uncover the critical dynamics at the depinning transition, and relate the
observed critical exponents to those of finite temperature creep dynamics.
Furthermore, we uncover the role of strength of the dipole-dipole interaction in determining the proper
dynamic universality class.

\vspace{3mm}


\noindent \textit{Model and method --}
We model the two-dimensional (2D) magnet subject to quenched disorder
by the 2D random-field Ising model with the dipole-dipole interaction, and
the dynamics is controlled by the external driving field:
\vspace{3mm}
\begin{equation}
H = -J \sum\limits_{\langle i,j\rangle }S_i S_j
    + V_{dd}\sum\limits_{i<j}\frac{S_i S_j}{r_{ij}^3}
    -\sum\limits_i (h_i+H)S_i,
\label{Hamiltonian}
\end{equation}
with $S_i=\pm 1$ at site $i$.
The first term of the Hamiltonian is the ferromagnetic coupling between one spin and its nearest neighbors.
Hereafter we measure the energy in the units of the coupling $J$.
The second term is the magnetic dipole-dipole interaction
with $r_{ij}=|i-j|$ and $V_{dd}$ a parameter for interaction strength.
The on-site random field $h_i$ distributes uniformly within an interval
$[-\Delta, \Delta]$ which generates random pinning potentials.
$H$ is a uniform magnetic field which drives the domain wall.

Our simulations are performed on $L \times L$ square lattice.
A flat domain wall between spin $+1$ and spin $-1$ is created
along $y$ axis at $x=1$ as the initial condition.
The magnetic field is applied to drive the domain wall in the positive
direction of $x$ axis in accordance to the Metropolis algorithm with single-spin flip\cite{MC}.
Periodic boundary condition (PBC) is adopted at the domain-wall ($y$) direction,
whereas Anti-periodic boundary condition (APBC)\cite{Nowak98} at the moving ($x$) direction.
The number of independent runs is at least $3000$.
The time unit is defined by a sweep of Monte Carlo trials over the whole system,
and the velocity is defined by $v=dM/2Ldt$ in steady states, with $M$ the total magnetization.
For a small system under large driving field, it reaches a steady state quickly
($t\sim10$ for warm-up and $\sim10^2$ for statistics),
whereas typically $10^3$ time steps for warm-up and statistics
with regard to large systems under critical driving field.
The time scale in MC technique should be proportional to the real time, but a
straightforward relation is not easy. In order to derive the correspondence, one need to compare the simulation
results and experiments at least once.

The system adopted in the present work is a coarse-grained one.
The on-site spin is represented in terms of a block spin which
contains $n_z \times n_{xy} \times n_{xy}$ unit cells
in a thin magnetic film, with $n_z$ and $n_{xy}$ the number of unit cells along the
out-of-plane and in-plane direction, respectively.
Then the energy unit $E_J = n_zn_{xy}aA$, with $a$ the lattice constant
of a real material and $A$ the corresponding exchange stiffness.
For $\textrm{Nd}_2\textrm{Fe}_{14}\textrm{B}$, $a = 0.88$ nm and $A = 7.7$ pJ/m\cite{Coey},
whereas $a = 0.25$ nm and $A = 10.3$ pJ/m for Co layer in Pt/Co/Pt thin films\cite{Coey,Metaxas07}.
The dimensionless temperature $T$ in the present work can be related to
the real temperature through: $T \cdot E_J /k_B$ , with $k_B$ the
Boltzmann constant. Taking $n_z = n_{xy} = 1$, one can have
$T = 0.1$ approximates to 49 K for $\textrm{Nd}_2\textrm{Fe}_{14}\textrm{B}$,
whereas 19 K for Pt/Co/Pt thin film.

\begin{figure} [t]
\centering
\includegraphics[width=.7\linewidth]{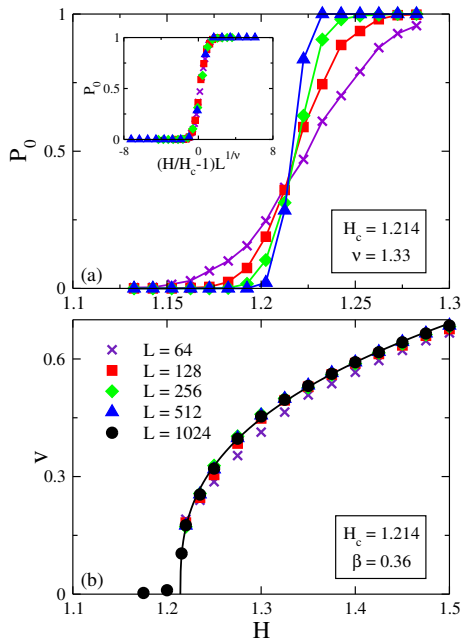}
\caption{(Color on line). (\textbf{a}):  The probability $P_0$ versus
the driving field with different system sizes and the corresponding scaling plot (inset).
$V_{dd}/J=0.1$ and $\Delta=1.5J$ are used all through this work.  (\textbf{b}): $v-H$ characteristics at zero temperature with difference system sizes.
The solid line is the fitting function (Eq. (\ref{velocity})) for $L=1024$.}
\label{fig1}
\end{figure}

\vspace{3mm}


\noindent \textit{Zero-temperature depinning --}
To come up with the quantitative description of depinning, we have to know its
key characteristic, the zero-temperature depinning field $H_c$.
Finding its true value is a challenge since in finite systems realizations of the
random potential fluctuate, and so do the corresponding values of the depinning field.

It is observed that for a given field and system size, the domain wall may either be pinned inside the sample, or it may go through from edge to edge.
We call the latter case a depinning event and evaluate the corresponding depinning probability $P_0$.
To determine a true value of $H_c$, one thus has to perform the finite-size
scaling analysis of $P_0$ which would contain $H_c$ as a parameter.
As shown in Fig.\,\ref{fig1}a, $P_0(L,H)$ increases sharply as function of the magnetic field
in the interval $H=1.1 \sim 1.3$.
The curves corresponding to different system sizes cross at point of $P_0=0.38\pm0.04$
at $H=1.214\pm0.006$.
This determines the depinning field which does not depend on the system size and thus can be taken as a
depinning field $H_c$ of a macroscopic system.

The problem of depinning at zero temperature is intimately related to percolation problem\cite{Vinokur:1993}.
We thus assume that the depinning probability function has the form characteristic to
the percolation problem\cite{Newman00}:
\begin{equation}
P_0(L,H)=\Phi[(H/H_c-1)L^{1/\nu}],
\label{p0}
\end{equation}
where $\nu$ is a universal exponent.
By choosing the variable $(H/H_c-1)L^{1/\nu}$ and $\nu$ as an adjustment parameter,
we find that at $\nu=1.33\pm0.05$,
all the data points of $P_0(L,H)$ collapse onto a single curve as shown in the inset of Fig.\,\ref{fig1}a.
This procedure defines the exponent $\nu$.

Now we are equipped to study the $v$-$H$ characteristics in the depinning regime.
We start with the zero-temperature behavior.
The results for $v$-$H$ for systems of different sizes are displayed in Fig.\,\ref{fig1}b.
One sees that for $L\geqslant 128$ the curves do not practically depend on the size of the system.
Assuming the standard $v$-$H$ depinning relation\cite{Fisher:1985}
\begin{equation}
v = v_0 (H/H_c-1)^\beta,
\label{velocity}
\end{equation}
where $\beta$ is a universal exponent, one finds
$v_0=1.16\pm0.01$, $H_c=1.214\pm0.006$ and $\beta = 0.36 \pm 0.01$ for $L=1024$.
This value is in a fair agreement with the $\beta=0.31$ result obtained in two-loops RG calculations\cite{LeDouss:2001}
showing that the elastic manifold approximation works pretty well at zero temperature.

\begin{figure*} [t]
\centering
\includegraphics[width=.75\linewidth]{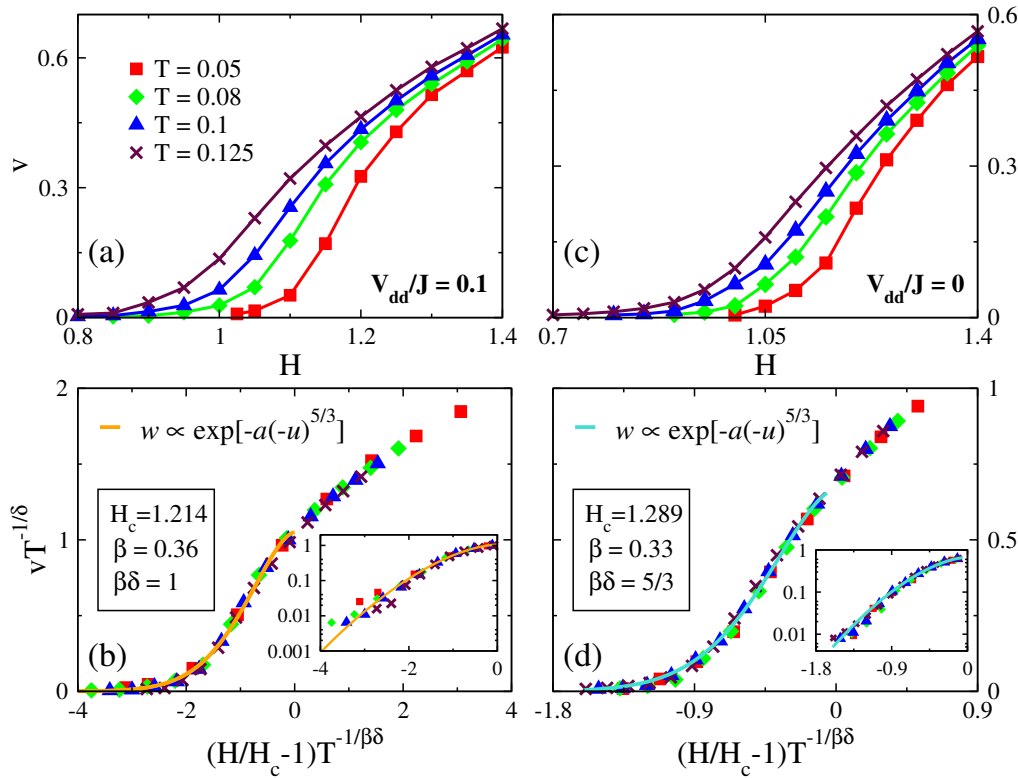}
\caption{(Color on line). (\textbf{a}):  Finite temperature $v-H$ characteristics for $V_{dd}/J = 0.1$.
(\textbf{b}): Scaling plot of $v-H$ curves as $vT^{-1/\delta}$ vs. $(H/H_c-1)T^{-1/\beta\delta}$.
The inset shows the same scaling behavior replotted in the semi-log scale.
(\textbf{c}):  Finite temperature $v-H$ characteristics for $V_{dd} = 0$.
(\textbf{d}): The corresponding scaling plot for $V_{dd} = 0$, the inset shows the same data in semi-log scale.}
\label{fig2}
\end{figure*}

\vspace{3mm}
\noindent \textit{Finite-temperature depinning --}
Now we turn to our main task, the finite temperature motion.
To reduce the computation time we choose $L=512$ system.
Figure\,\ref{fig2}a shows the expected increase in velocity at the given field upon increasing temperature
and an appreciable tail below the depinning field due to thermally activation processes.

We use the standard scaling ansatz\cite{Nowak98,Nowak99,Hu07}:

\begin{equation}
v(T,H) = T^{1/\delta}\Psi\left[(H/H_c-1)T^{-1/\beta\delta}\right],
\label{th-ansatz}
\end{equation}
with $\Psi(x)\sim x^\beta$ as $x\rightarrow\infty$.
We achieve the best collapse of the data to a single curve
with $\delta = 2.76\pm0.02$ by adopting the values of $H_c$ and $\beta$ determined above,
see Fig.\,\ref{fig2}b.
Note, that these results cease to hold for large dd interaction, $V_{dd}/J>0.5 $
where the ferromagnetic order is broken.

At $x<0$ the scaling function exhibits the asymptotic behavior $\Psi(x)=1.05\exp\left[-0.70(-x)^{5/3}\right]$, see
 Fig.\,\ref{fig2}b, and one arrives at the dynamics of the domain wall across the transition\cite{Nowak98}
given by:

\begin{equation}
v = v_1T^{1/\delta}\exp\left\{-\left[\frac{E_c}{T}\left(1-\frac{H}{H_c}\right)\right]^{5/3}\right\},
\label{th-function}
\end{equation}
where $E_c\approx0.81$ is an energy barrier which governs the domain-wall velocity at finite temperatures,
and the condition $\beta\delta=1$ is taken into account.
Notably, the domain-wall motion is not the conventional Arrhenius-type.
The origin of this  nontrivial temperature dependence is the renormalization of the random potential landscape by thermal
fluctuations.
\begin{figure*}[t]
\centering
\includegraphics[width=0.8\linewidth]{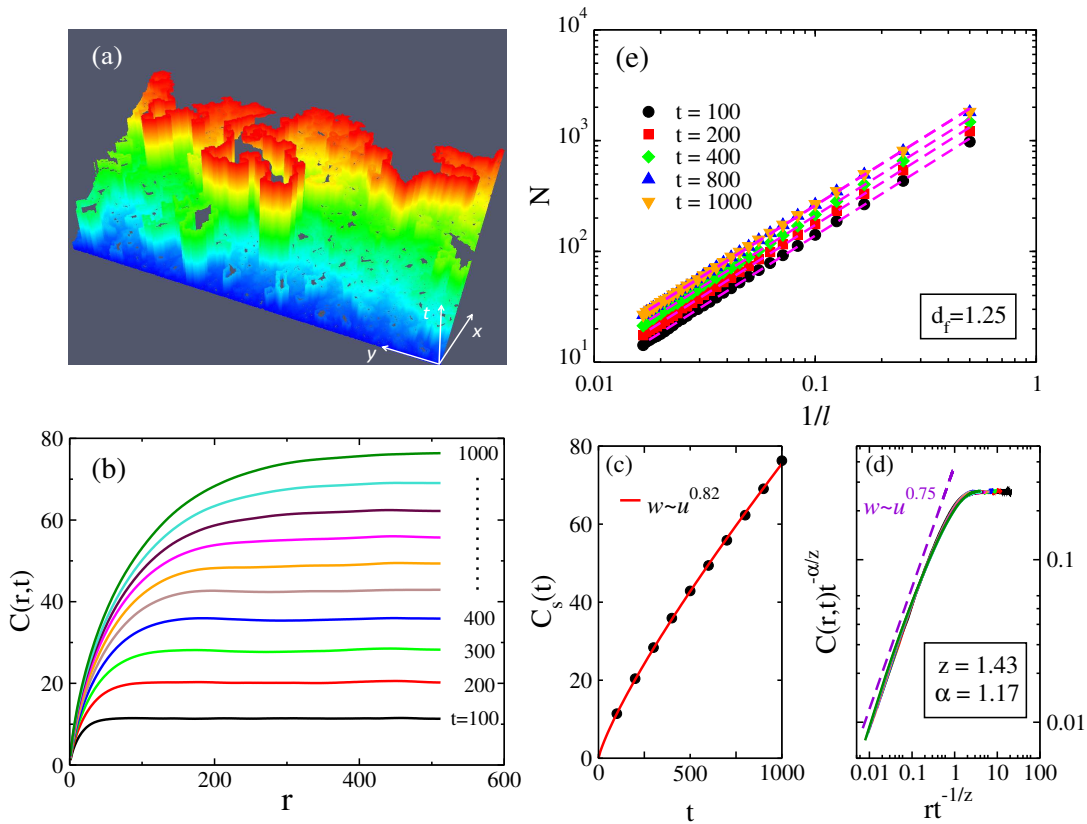}
\caption{(Color on line). (\textbf{a}): Time evolution of domain wall near the depinning threshold at zero temperature.
(\textbf{b}): Height-difference correlation function $C(r,t)$ versus $r$ for different values of $t$ near the depinning threshold.
(\textbf{c}): Saturated value $C_s(t)$ versus $t$. (\textbf{d}): Scaling plot for data in (\textbf{b}) with the dashed line $y\sim x^{2-d_f}$.
(\textbf{e}): Log-log plot of $N$ versus $1/l$ at different times, with $l$ the ruler size and $N$ the measured length in units of $l$.
 Parameters used in calculations are: $L=1024$, $V_{dd}/J=0.1$ and $H=1.22$.}

\label{fig3}
\end{figure*}

The exponent 5/3 is in an excellent agreement with the prediction of Ref. \cite{Ioffe:1987} (see Ref. \cite{FV1990}
for thermal depinning of vortex systems).  This establishes an intimate connection between
the critical depinning behavior of the domain wall at finite temperatures and thermal softening of the pinning potential.
Furthermore, juxtaposing our $v$-$H$ curves with those obtained earlier
for the case $V_{dd}=0$\cite{Nowak98}, one makes a remarkable observation. Figure\,\ref{fig2}c shows the $v(H)$ dependencies that
at the first glance are not that different from those of Fig.\,\ref{fig2}a.
The scaling treatment, however, yields $v\propto\exp\left[-U_c(1-H/H_c)^{5/3}/T\right]$, see Fig.\,\ref{fig2}d,
i.e. the Arrhenius activation behaviour with the barrier that scales as  $U_c(1-H/H_c)^{5/3}$,
where $U_c \approx 0.67$ is the bare energy barrier, $H_c=1.289$, $\beta=0.33$, and $\beta\delta=5/3$.
We now recall that this kind of thermally activated behavior at $V_{dd}=0$ is expected
for the case of the motion of DW in the field of the random force\cite{Ioffe:1987},
where an impurity ``knows" to which magnetic domain it belongs in.
This implies that the pinning energy barrier comprises the energies of all
pinning sites located in the area spanned by the domain wall
during an elemental activation jump. In this case the contribution
of the fluctuation thermal broadening of the domain wall position
gives the negligible contribution into the total pinning energy
thus thermal fluctuations cannot appreciably reduce the depinning field.
Hence the thermally activated motion retains its Arrhenius-like character.
We conclude that varying $V_{dd}$ one tunes the system between the
random force and random potential pinning behaviors.
Hence our findings provide an irreplaceable tool for identifying these distinct pinning
mechanisms in the experiment.
The behaviors summarized in Fig.\,\ref{fig2} constitute the main results of our work.

Another comment in order is that at finite temperatures the definition of the depinning field
$H_c$ is not straightforward. Our approach offers a systematic way for analyzing data
at finite temperatures yielding $H_c$ and the thermal activation energy barrier $E_c$ simultaneously.
Importantly, $E_c$ depends not only on the strength of randomness $\Delta$ but also
on the competition of the exchange coupling $J$
and the dipole-dipole interaction strength $V_{dd}$.


\vspace{3mm}
\noindent \textit{Domain-wall morphology --}
Next we investigate the DW morphology during the depinning process for the system with $V_{dd}/J=0.1$.
To this end we set a flat domain wall along $y$ axis at $x=1$ at $t=0$ with $H=1.22$
just above $H_C=1.214$, and drive it along $x$ direction at zero temperature.
As shown in Fig.\,\ref{fig3}a, the domain wall evolves rougher with time,
and develops a fractal structure. Moreover, there remain several
small unflipped-spin areas (black puddles) forming the `` lakes'' inside the domain.
The multiconnected nature of the flipped domain originates from spatial
fluctuations of the depinning field due to random character of the pinning potential:
there are lacoons where the local depinning field still exceeds the driving field.
The ``overhangs" of the frontier of the domain wall are of the same origin.
To quantify this multi-valued domain-wall morphology, we define the modified height function $h(y,t)$
\cite{Meakin}
\begin{equation}
h(y,t)=\sum_{x=1}^L \theta[S_{x,y}(t)],
\label{heightfunction}
\end{equation}
with $\theta(x)$ the unit step function and $S_{x,y}(t)$ the spin value on the site (x,y) at time $t$.
The function $h(y,t)$ describes the total number of flipped spins along the line $y$ at time $t$.
It is obvious that $h(y,t)$ describes the position of domain wall if there is no `` lakes'' and `` overhangs''.

The height-difference correlation function $C(r,t)$ describes the domain-wall roughness characteristics
and is define as\cite{Meakin,Kim89}:
\begin{equation}
C(r,t)=\sqrt{\left\langle\left[h(y+r,t)-h(y,t)\right]^2 \right\rangle},
\label{correlationfunction}
\end{equation}
with $r$ the distance between two points in $y$ direction.

As shown in Fig.\,\ref{fig3}b, for fixed $t$, $C(r)$ increases from zero with $r$ and saturates
at large $r$, and the saturated value $C_s(t)$ increases with $t$.
These properties can be understood from the time-evolution process of domain-wall morphology as
displayed in Fig.\,\ref{fig3}a.
The initial domain wall is a straight blue line with no height difference.
By applying driving field, locally meandering segments appear first (see the light blue regions in Fig.\,\ref{fig3}a).
Correlation of domain-wall positions only exists in a small length scale.
As time evolves, the meandering segments spread out along both the domain-wall ($\parallel$) and moving ($\perp$) directions,
leading to rougher structures.
There are two $t$-dependent correlation lengths:
$\xi_\parallel(t)$ and $\xi_\perp(t)$,
growing with time as
$\xi_\perp(t)\sim\xi_\parallel(t)^{\alpha}\sim t^{\alpha/z}$
with $z$ the dynamic exponent and $\alpha$ the roughness exponent,
and the correlation function evolves as\cite{Meakin}:
\begin{equation}
C(r,t)\sim\xi_\perp(t)g[r/\xi_\parallel(t)]\sim t^{\alpha/z}g(r/t^{1/z}),
\label{crt-ansatz}
\end{equation}
with $g(x)$ saturates at constant as $x>>1$.

As displayed in Fig.\,\ref{fig3}c, we obtain $\alpha/z=0.82\pm0.01$ in terms of $C_s(t)\sim t^{\alpha/z}$
in the large $r$ limit of Eq.\,(\ref{crt-ansatz}).
By choosing $z=1.43\pm0.01$, all the data collapse into a single curve as displayed in
Fig.\,\ref{fig3}d, which determines the dynamics exponent $z$.
The roughness exponent $\alpha$ is then estimated as $\alpha=1.17\pm0.02$.

We then study the fractal geometry of the domain wall.
For a fractal structure, the measured length $N$ in units of ruler size is related to the ruler size $l$ by:
$d_f=\log N/\log(1/l),$
with $d_f$ the fractal index.
Through log-log plot of $N$ versus $1/l$ as shown in Fig.\,\ref{fig3}e,
we obtain $d_f=1.25\pm0.01$. We notice that $C(r,t)\sim r^{D-d_f}$ only holds for $r<\xi_\parallel(t)$\cite{Meakin}.
As shown by the dashed line in Fig.\,\ref{fig3}d, the exponent $2-d_f\approx 0.75$ appears in the small scaling variable $r/t^{1/z}$
limit of function $g(r/t^{1/z})$.
Importantly, the fractal morphology is a signature of the depinning region.
At large drives $H$, the DW gets flat\cite{Nattermann:2000}.

\vspace{3mm}
\noindent \textit{Discussions and conclusions --}
To conclude, we have investigated the depinning dynamics
of magnetic domain wall with the dipole-dipole interaction,
and obtained five critical exponents.
It is confirmed that the scaling relation
$\beta/\nu=z-\alpha$\cite{Nattermann:2000} is satisfied.
The case without the dipole-dipole interaction has been investigated previously by
MC simulations\cite{Nowak98} and Langevin dynamics\cite{Rosso03,Duemmer05}.
The critical exponents are different for the two cases, indicating the two cases
belong to different universality classes.

The existence of two universal classes in elastic manifolds in random potentials has
been addressed by two of the present authors\cite{Hu07} for vortex dynamics
in type-II superconductors, where Bragg glass and amorphous
vortex glass (AVG) correspond to weak and strong random pinning potentials.
In heavily disordered AVG the depinning dynamics is of Arrhenius-type, while in the ordered Bragg
glass state one observes non-Arrhenius-type behavior, similar to the present domain-wall system.
 Interestingly, scaling functions of two universality classes in vortex
dynamics share a simple exponential form, which differs from our results for
the different dimensionality of the space and the elastic manifold;
this intriguing issue calls for further investigation.

\noindent \textit{Acknowledgements --}
We are delighted to thank Andreas Glatz for useful discussion and critical reading of the manuscript.
This work was supported by  the WPI initiative
on Materials Nanoarchitectonics, MEXT of Japan (BX, MBL, and XH) and the Elements Strategy Initiative Center for
Magnetic Materials under the outsourcing project of MEXT, and by the U.S.
Department of Energy, Office of Science, Materials Sciences and Engineering
Division (VV).


\end{document}